\documentclass{article}
\usepackage{fullpage}
\usepackage{amssymb, amsmath, amsthm}
\usepackage{graphicx}
\usepackage{xspace}
\usepackage{todonotes}
\usepackage[all=normal,bibliography=tight]{savetrees}
\usepackage{enumerate}

\newtheorem{theorem}{Theorem}

\newtheorem{claim}{Claim}

\newcommand{\cqed}{\renewcommand{\qedsymbol}{$\lrcorner$}}

\newcommand{\threepartition}{\textsc{3-Partition}\xspace}
\newcommand{\strippacking}{\textsc{Strip Packing}\xspace}
\title{Hardness of approximation for strip packing}

\author{Anna Adamaszek\thanks{University of Copenhagen, Denmark, \texttt{anad@di.ku.dk}. Research supported by the Danish Council for Independent Research DFF-MOBILEX mobility grant.}
   \and Tomasz Kociumaka\thanks{University of Warsaw, Poland, \texttt{kociumaka@mimuw.edu.pl}.}
   \and Marcin Pilipczuk\thanks{University of Warsaw, Poland, \texttt{malcin@mimuw.edu.pl}. Research supported by Polish National Science Centre grant UMO-2013/09/B/ST6/03136.}
   \and Micha\l{} Pilipczuk\thanks{University of Warsaw, Poland, \texttt{michal.pilipczuk@mimuw.edu.pl}. Research supported by Polish National Science Centre grant UMO-2013/11/D/ST6/03073. Author
   supported by Foundation for Polish Science (FNP) via the START stipend programme.}
}
\date{}

\begin{document}
\maketitle
\begin{abstract}
Strip packing is a classical packing problem, where the goal is to pack a set of rectangular objects into a strip of a given width, while minimizing the total height of the packing. The problem has multiple applications, e.g. in scheduling and stock-cutting, and has been studied extensively.

When the dimensions of objects are allowed to be exponential in the total input size, it is known that the problem cannot be approximated within a factor better than $3/2$, unless $\mathrm{P}=\mathrm{NP}$. However, there was no corresponding lower bound for polynomially bounded input data. In fact, Nadiradze and Wiese [SODA 2016] have recently proposed a $(1.4 + \epsilon)$ approximation algorithm for this variant, thus showing that strip packing with polynomially bounded data can be approximated better than when exponentially large values in the input data are allowed. Their result has subsequently been improved to a $(4/3 + \epsilon)$ approximation by two independent research groups [FSTTCS 2016, arXiv:1610.04430]. This raises a question whether strip packing with polynomially bounded input data admits a quasi-polynomial time approximation scheme, as is the case for related two-dimensional packing problems like maximum independent set of rectangles or two-dimensional knapsack.

In this paper we answer this question in negative by proving that it is NP-hard to approximate strip packing within a factor better than $12/11$, even when admitting only polynomially bounded input data. In particular, this shows that the strip packing problem admits no quasi-polynomial time approximation scheme, unless $\mathrm{NP} \subseteq \mathrm{DTIME}(2^{\mathrm{polylog}(n)})$.
\end{abstract}

\section{Introduction}

In the strip packing problem we are given a collection of rectangular items $I$ where each item $i \in I$ is defined by its width $w_i \in \mathbb{N}$ and height $h_i\in \mathbb{N}$, together with a width parameter $W \in \mathbb{N}$. The goal is to pack all items of $I$ into a strip of width $W$ such that no two rectangles overlap, and the height of the packing is minimized. Formally, in a packing each item $i \in I$ corresponds to a rectangle $(x_i, x_i + w_i) \times (y_i, y_i + h_i)$, where $x_i, y_i \in \mathbb{N}$ and $x_i + w_i \le W$, and the corresponding (open) rectangles are pairwise disjoint. The objective is to construct a packing (i.e., find values of $x_i, y_i$ for all the input rectangles) minimizing the value of $H = \max_{i \in I} (y_i + h_i)$.

\paragraph{Motivation.}
The strip packing problem arises naturally in many settings.
In the area of scheduling, it models scheduling of jobs where each job $i$ requires a contiguous portion of $w_i$ memory, where the total memory of the machine is $W$, over a period of time $h_i$. Minimizing the total height  $H$ of the solution corresponds to minimizing the makespan of the schedule. In the area of industrial manufacturing, we have the following cutting stock problem which also corresponds to strip packing. We want to cut rectangular pieces out of a sheet of material (e.g. cloth or wood) of fixed width, minimizing the total amount of material used. Here we do not allow the rectangles to be rotated (because of the constraints like patterns on the material or the grain of the wood). Recently, strip packing has also been applied in electricity allocation and peak demand reductions in smart-grid (see \cite{GGIK16}).

\paragraph{Related work.}
The strip packing problem has been studied extensively since 1980, when Baker et al. \cite{BCR80} provided an algorithm with an asymptotic approximation ratio of $3$. Subsequently, in a series of papers \cite{CGJT80, Sleator80, Schiermeyer94, Steinberg97, HS09, HJPS14} better approximation algorithms have been presented. Currently, the best result is a $5/3+\epsilon$ approximation ratio by Harren et al. \cite{HJPS14}. The best known lower bound for approximating the problem is $3/2$, and it can be shown by a straightforward reduction from \textsc{Partition}.

The asymptotic approximation ratio for strip packing has also been studied in \cite{CGJT80, Golan81, BBK81}. There, the lower bound of $3/2$ does not hold. In fact, there is an asymptotic FPTAS given by Kenyon and R{\'{e}}mila \cite{KR00} with an additive constant of $O(h_{\textrm{max}}/\epsilon^2)$, and an asymptotic PTAS given by Jansen and Solis{-}Oba \cite{JS09} with an additive constant of $h_{\textrm{max}}$, where $h_{\textrm{max}}$ is the maximum height of a rectangle in the input instance. Strip packing has also been studied in the setting where rotations of the items by $90$ degrees are allowed. In this setting, Jansen and van Stee \cite{JS05} provided an asymptotic FPTAS.

Recently, pseudo-polynomial time algorithms for strip packing have been considered. Nadiradze and Wiese \cite{NW16} have given an algorithm achieving an approximation ratio of $1.4+\epsilon$, which has then been refined to a $4/3+\epsilon$-approximation algorithm by G{\'{a}}lvez et al. \cite{GGIK16} and by Jansen and Rau \cite{JR16}. The running time of these algorithms is polynomial when input data is polynomially bounded, that is, 
when all the numbers $W$, $w_i$ and $h_i$ are bounded polynomially in the number of items.

Strip packing is related to the \emph{geometric knapsack} problem. There, given a collection of rectangular items $I$, where each item $i \in I$ is defined by its width $w_i \in \mathbb{N}$, height $h_i\in \mathbb{N}$, and weight $\omega_i\in \mathbb{R}^+$, and a rectangular box of size $W \times H$, the goal is to pack a subcollection of items of maximum total weight into the given box so that no two items overlap. For this problem there is a $(2+\epsilon)$-approximation algorithm by Jansen and Zhang \cite{JZ07}, and a QPTAS by Adamaszek and Wiese \cite{AW15}. 

\paragraph{Our results.}
The QPTAS for geometric knapsack \cite{AW15}, and also the pseudo-polynomial time algorithms for strip packing \cite{NW16, GGIK16, JR16}, are based on the machinery introduced by Adamaszek and Wiese \cite{AW13} for designing a quasi-polynomial time approximation scheme for maximum independent set of rectangles. This raises a question, asked in~\cite{GGIK16, JR16}, whether there is also a quasi-polynomial time approximation scheme for strip packing, possibly based on the same machinery, when the input data is polynomially bounded.

We give a negative answer to this question, by showing APX-hardness of strip packing. This shows that strip packing behaves differently with respect to approximation than the related maximum independent set of rectangles and two-dimensional knapsack problems.

\begin{theorem}\label{thm:hardness}
For every $\varepsilon > 0$, it is NP-hard
to approximate \strippacking{} within a factor of $12/11-\varepsilon$,
   even if the dimensions of the rectangles are given in unary.
\end{theorem}

At the heart of our reduction in the proof of Theorem~\ref{thm:hardness} lies an example showing that a rearrangement argument, being the core engine of the approach of Nadiradze and Wiese~\cite{NW16}
(cf. Section~4 of~\cite{NW16}), fails to work if there are three rows of rectangles, not two as in~\cite{NW16}.
Figure~\ref{fig:packing} illustrates this example: if we pick $a$ and $b$ to be such integers that any nontrivial integral solution to the equation $ax+by+z=0$ requires integers of magnitude
much larger than the number of rectangles, then one can argue that the presented packing in a strip of height $11$ is essentially the only one possible, and any rearrangement requires a strip of height of at least $12$. Our reduction exploits this figure by chopping every $b \times 1$ rectangle vertically into three pieces, embedding a \threepartition instance into the picture.

At the intuitive level, it is the combination of hard constraints---the requirement of packing all the items and the inability of widening the strip---that makes the problem hard to approximate.
The issue lies in relatively tall items that have to ``stick out'' by a significant portion of their height in case they cannot be packed optimally into some prescribed space.
In fact, Nadiradze and Wiese~\cite{NW16} have given a PTAS for the case when a constant (depending on $\varepsilon$) number of items can be dropped, even when the input data is exponentially bounded.

\newcommand{\HAll}{11}
\newcommand{\HAllPlus}{12}
\newcommand{\HTopA}{4}
\newcommand{\HTopB}{5}
\newcommand{\HMidA}{1}
\newcommand{\HMidB}{2}
\newcommand{\HMidC}{3}
\newcommand{\R}{\mathcal{R}}

\section{Hardness of approximation for strip packing}

In this section, we prove Theorem~\ref{thm:hardness}
by a reduction from the \threepartition problem,
   which is known to be strongly NP-complete~\cite{GJ79}.

Assume we are given an instance of the \threepartition problem:
a multiset $S = \{s_1,s_2,\ldots, s_{3n}\}$ of $3n$ integers, summing up to zero; the goal is to partition $S$ into $n$ triples,
each summing up to zero.
Furthermore, let $M = 1 + \sum_{i=1}^{3n} |s_i|$.
Since \threepartition is strongly NP-hard, we can assume
that all integers $s_i$ are given in unary; that is,
the running time bound, the dimensions of the rectangles, and the size of the output instance
in our reduction can depend polynomially on both $n$ and $M$.

\paragraph{Construction.}
We start by choosing two positive integers $a,b$ using
the following standard claim.
\begin{claim}\label{cl:pick-abc}
Given positive integers $n$ and $M$, one can in time polynomial in $n+M$
find two positive integers $a$, $b$
with the following properties:
\begin{enumerate}[(I)]
\item\label{p:lower} $a,b > 3M$ and $b$ is divisible by $3$;
\item\label{p:dioph} for every three integers $x,y,z$ with
$|x|,|y|,|z| \le \max(9n,3M)$, if $ax+by+z= 0$, then $x=y=z=0$;
\item\label{p:upper} $a$ and $b$ are bounded polynomially in $n$ and $M$.
\end{enumerate}
\end{claim}

\begin{proof}
Let us take $b=\max(9n,3M)+3$ and $a=b^2$.
Let $x,y,z,w$ be integers such that $ax+by+z=0$ and $|x|,|y|,|z|\le \max(9n,3M)$.
Consequently, we have $xb^2+yb+z=0$ and $|x|,|y|,|z|<b$.
Taking the equality modulo $b$, we see that $z$ is divisible by $b$ and thus
$z=0$. Then, analyzing it modulo $b^2$, we show that $y$ is divisible by $b$
and hence $y=0$. Finally, we conclude that $x=0$. \cqed\end{proof}

Armed with Claim~\ref{cl:pick-abc}, we can now construct
rectangles in the output \strippacking instance.
First, we set the width of the strip to
\begin{equation}
W := 2(a+b) \cdot n.
\end{equation}
We will now define a set of rectangles of total area
$\HAll W$, such that the rectangles can be packed into a rectangle
$W \times \HAll$ if and only if the input \threepartition instance
is a yes-instance. As every rectangle in our construction will
have integral height, if the input \threepartition instance
is a no-instance, then we will need a strip of width at least
$\HAllPlus$ to accommodate all rectangles. Consequently, such a construction
would prove Theorem~\ref{thm:hardness} due to (strong)
NP-completeness of \threepartition.

The output instance consists of the following rectangles:
\begin{description}
\item[(middle rectangles)]
We construct in total $6n$ \emph{middle rectangles} as follows:
\begin{itemize}
\item $2n$ rectangles of height $\HMidB$ and width $a$;
\item $n$ rectangles of height $\HMidC$ and width $b$;
\item for every $1 \leq i \leq 3n$, a rectangle
of height $\HMidA$ and width $b/3 + s_i$ (called \emph{solution rectangles}).
\end{itemize}
\item[(side rectangles)]
We construct in total $4n+1$ \emph{side rectangles} as follows:
\begin{itemize}
\item $2n$ rectangles of height $\HTopA$ and width $a+b$;
\item $2n-1$ rectangles of height $\HTopB$ and width $a+b$;
\item one rectangle of height $\HTopB$ and width $a$, and one
rectangle of height $\HTopB$ and width $b$.
\end{itemize}
\end{description}
Note that $b> 3M$ implies that all $b/3>s_i$ for $1\le i\le 3n$ and thus all rectangles are well-defined.

\begin{figure}[htbp]
\begin{center}
\includegraphics{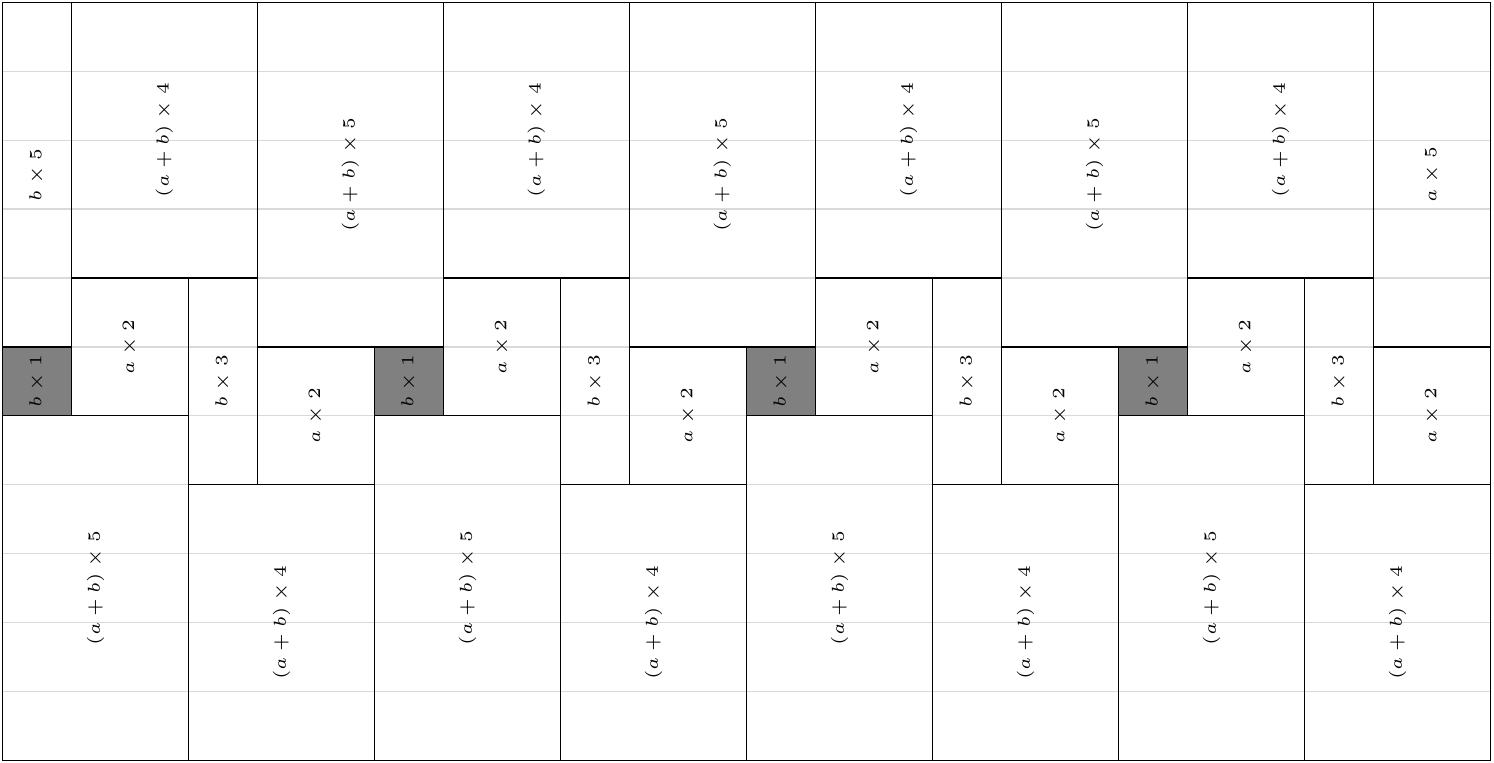}
\end{center}
\caption{Canonical packing of the rectangles. Every gray $b \times \HMidA$ rectangle consists of three solution rectangles,
  corresponding to a set in the solution to the input \threepartition instance.}\label{fig:packing}
\end{figure}

\paragraph{From partition to packing.}
We now show that if the input \threepartition instance is a yes-instance, then we can pack all rectangles into a strip of size $W \times \HAll$.
Let us group solution rectangles into $n$ triples, as in the solution to the input \threepartition instance.
In this manner, we replace $3n$ solution rectangles with $n$ middle rectangles, each of width $b$ and height $\HMidA$.
Such rectangles can be arranged as in Figure~\ref{fig:packing}:
\begin{enumerate}
\item We put $n$ side rectangles of dimension $(a+b) \times \HTopA$ and $n$ side rectangles of dimension $(a+b) \times \HTopB$ on the bottom of the $W \times \HAll$ strip
in an alternating fashion, starting from a rectangle of height $\HTopB$.
\item We put $n$ side rectangles of dimension $(a+b) \times \HTopA$, $n-1$ side rectangles of dimension $(a+b) \times \HTopB$, and the two remaining side rectangles of dimension $a \times \HTopB$ and $b \times \HTopB$
on the top of the $W \times \HAll$ strip, in the following order: first the $b \times \HTopB$ rectangle, then an alternating sequence of $(a+b) \times \HTopA$ and $(a+b) \times \HTopB$ rectangles, starting
from an $(a+b) \times \HTopA$ rectangle, and the $a \times \HTopB$ rectangle at the end.
\item In the middle, we put the middle rectangles, in the following order of heights: $\HMidA, \HMidB,\HMidC,\HMidB,\ldots,\HMidA, \HMidB,\HMidC,\HMidB$.
\end{enumerate}
A direct check shows that the middle rectangles fit exactly in the spaces left by the side rectangles.

\paragraph{From packing to partition.}
Assume now that all rectangles can be packed into a strip of size $W \times \HAll$; let us fix such a packing. Our goal is to show that the input \threepartition instance is a yes-instance,
and the route to proving this is by verifying that the arrangement of non-solution rectangles must be exactly as depicted in Figure~\ref{fig:packing}.
A direct calculation shows that the total area of all rectangles is $\HAll W$. Consequently, the $W \times \HAll$ strip is completely covered by the rectangles.

We say that a vertical line is \emph{in general position} if it intersects the strip but does not contain any side of a rectangle.
We say that a set $\R$ of rectangles is \emph{tightly packed} if every vertical line in general position intersects
exactly one rectangle of $\R$. In other words, $\R$ is tightly packed if there is exactly one rectangle of $\R$ touching the left side of the strip, exactly one touching the right side,
and when we scan the strip from left to right, then a rectangle from $\R$ ends at some $x$-coordinate within the strip if and only if a new one rectangle starts at exactly the same $x$-coordinate.

We now show the following.
\begin{claim}\label{cl:side-mid-side}
Every vertical line in general position intersects exactly three rectangles: two side ones and one middle one.
Consequently, the set of middle rectangles is tightly packed, and one can partition the set of side rectangles into two parts $\R_1$ and $\R_2$ such that $\R_t$ is tightly packed for $t=1,2$.
\end{claim}
\begin{proof}
Since every rectangle has a height of at most $\HTopB$, any two rectangles have a total height of at most $10$, and thus every vertical line in general position intersects at least three rectangles.
Since the total width of all rectangles is exactly $3W$, every vertical line in general position intersects exactly three rectangles. 
However, since any three side rectangles have a total height of at least $12$, every vertical line in general position can intersect only two of them.
As the side rectangles have a total width of
exactly $2W$, every vertical line in general position intersects exactly two side rectangles, and thus exactly one middle rectangle. 

For the second claim, we can construct $\R_1$ by taking one of the two side rectangles at the left side of the strip, and build $\R_1$ from left to right by picking
a side rectangle starting at the $x$-coordinate where the previously picked side rectangle ends. There will be always such a side rectangle, because the vertical line in general position just after this
$x$-coordinate again intersects two side rectangles.
\cqed\end{proof}

Let $\R_1$ and $\R_2$ be as in Claim~\ref{cl:side-mid-side}. Observe that, in each set $\R_t$, the total width of all rectangles equals $W = 2(a+b)n$.
Since there are only $4n+1$ side rectangles, from Claim~\ref{cl:pick-abc}, property~\eqref{p:dioph}, it easily follows that one of the sets $\R_t$ (w.l.o.g., say it is $\R_1$), contains $n$ rectangles $(a+b) \times \HTopA$
and $n$ rectangles $(a+b) \times \HTopB$, and the other one contains the remaining $2n+1$ rectangles, including the two rectangles $a \times \HTopB$ and $b \times \HTopB$.
By potentially taking a symmetrical image of the strip, we may assume that the $b \times \HTopB$ rectangle appears to the left of the $a \times \HTopB$ rectangle in the strip.

Assume that the left end of the strip is at $x$-coordinate $0$, and the right end is at $x$-coordinate $W$.
\begin{claim}\label{cl:x-coor}
Let $x_0$ be an $x$-coordinate where a side rectangle from $\R_t$ ($t \in \{1,2\}$) ends. Then $x_0 = a \cdot n_a + b \cdot n_b$
for some integers $0 \leq n_a,n_b \leq 2n$ with $n_b - n_a \in \{0,1\}$.
\end{claim}
\begin{proof}
Since every side rectangle has width $a+b$, $a$, or $b$, and the rectangles of $\R_t$ are tightly packed, we can express $x_0 = (a+b) m_{a+b} + a m_a + c m_b$, where $m_z$ is the number of rectangles of width $z$ that are in $\R_t$ and to the left of $x_0$.
By taking $n_a = m_{a+b} + m_a$ and $n_b = m_{a+b}+m_b$, we obtain $x_0 = a \cdot n_a + b \cdot n_b$ and $0 \leq n_a,n_b \leq 2n$.

Furthermore, $n_b- n_a = m_b - m_a$. Since there is only one rectangle of width $b$ and one of width $a$, we have $m_a,m_b \in \{0,1\}$. Since we have assumed that the $b \times \HTopB$ one is to the left of the $a \times \HTopB$ one, it cannot hold that $m_b = 0$ and $m_a = 1$.
Consequently, $m_b - m_a \in \{0,1\}$.
\cqed\end{proof}

\begin{claim}\label{cl:middle-batch}
If we order the middle rectangles from left to right (recall that they are tightly packed), then 
every maximal consecutive segment of solution rectangles consists of three
rectangles of total width exactly $b$.
\end{claim}
\begin{proof}
Let $R_{i_1},R_{i_2},\ldots,R_{i_k}$ be a maximal consecutive segment of solution rectangles in the left-to-right ordering of all middle rectangles.
Assume that the width of $R_{i_j}$ is $b/3+s_{i_j}$.
We shall prove that $k=3$ and that $\sum_{j=1}^k s_{i_j}=0$.

Let $x_0$ be the $x$-coordinate of the left side of $R_{i_1}$ and $x_0'$ be the $x$-coordinate of the right side of $R_{i_k}$.
Since the sequence $R_{i_1},R_{i_2},\ldots,R_{i_k}$ is maximal, other middle rectangles have different heights, and the whole $W \times \HAll$ strip is covered by rectangles, both at $x_0$ and at $x_0'$ a side rectangle ends. By Claim~\ref{cl:x-coor} we have
$x_0 = an_a + bn_b$ and $x_0' = an_a' + bn_b'$ for some integers $n_a,n_b,n_a',n_b'$
with $\delta := n_b-n_a \in \{0,1\}$ and $\delta' := n_b'-n_a' \in \{0,1\}$.
Furthermore, $x_0' - x_0 = kb/3 + \sum_{j=1}^k s_{i_j}$, that is,
$$kb/3 + \sum_{j=1}^k s_{i_j} = x_0' - x_0 = a(n_a'-n_a) + b(n_b'-n_b),$$
i.e.,
$$0 = a(3n_a'-3n_a) + b(3n_b'-3n_b-k)-3\sum_{j=1}^k s_{i_j}.$$
Note that $|3n_a'-3n_a|\le 6n\le\max(9n,3M)$, $|3n_b'-3n_b-k|\le 9n\le\max(9n,3M)$, and $|3\sum_{j=1}^k s_{i_j}|\le3M\le\max(9n,3M)$.
Thus, by Claim~\ref{cl:pick-abc}, property~\eqref{p:dioph}, we easily obtain $n_a' = n_a$, $n_b'-n_b=k/3$, and $\sum_{j=1}^k s_{i_j}=0$. 
However, recall that $\delta = n_b-n_a \in \{0,1\}$ and $\delta' = n_b'-n_a' \in \{0,1\}$ while $k \geq 1$.
Then $\delta'=\delta+k/3$, so $\delta,\delta'\in \{0,1\}$ and $k\geq 1$ implies that the only possibility is $k=3$, $\delta = 0$, and $\delta' = 1$.
\cqed\end{proof}
Consequently, the solution rectangles are partitioned into triples of rectangles of total width $b$.
Such a partition of the solution rectangles induces a solution to the input \threepartition instance.
This finishes the proof of Theorem~\ref{thm:hardness}.

\bibliographystyle{plain}
\bibliography{bib}


\end{document}